\documentclass[twocolumn]{aastex7}

\def\lapp{\ifmmode\stackrel{<}{_{\sim}}\else$\stackrel{<}{_{\sim}}$\fi}
\def\gapp{\ifmmode\stackrel{>}{_{\sim}}\else$\stackrel{>}{_{\sim}}$\fi}
\usepackage{graphicx}

\usepackage{multirow}
\usepackage{color}
\usepackage{amsmath}
\usepackage{soul}
\usepackage{hyperref}
\usepackage{lineno}
\usepackage{mathrsfs}
\linenumbers
\newcommand{\fluxcgs}{\ensuremath{\mathrm{erg}\,\mathrm{s}^{-1}\,\mathrm{cm}^{-2}}}

\def\footnoterule{\hrule width 0.3\textwidth \vspace*{0.1cm}}

\shorttitle{}
\shortauthors{}
\begin{document}

\title{A Comprehensive X-ray Characterization of the Pulsar Wind Nebula CTB 87: Evidence for Reverse-Shock Interaction from Its X-ray and Broadband Emissions}

\correspondingauthor{Hongjun An}
\email{hjan@cbnu.ac.kr}

\author[0000-0002-9103-506X]{Jaegeun Park}
\email{}
\affiliation{Department of Astronomy and Space Science, Chungbuk National University, Cheongju, 28644, Republic of Korea}
\affiliation{Columbia Astrophysics Laboratory, 550 West 120th Street, New York, NY 10027, USA}
\author[orcid=0000-0002-2312-8539]{Jason Alford}
\email{}
\affiliation{New York University Abu Dhabi, PO Box 129188, Abu Dhabi, UAE}
\author[0000-0002-6389-9012]{Hongjun An}
\email{}
\affiliation{Department of Astronomy and Space Science, Chungbuk National University, Cheongju, 28644, Republic of Korea}
\author[orcid=0000-0003-4679-1058]{Joseph D. Gelfand}
\email{}
\affiliation{New York University Abu Dhabi, PO Box 129188, Abu Dhabi, UAE}
\author[0000-0003-0226-9524]{Chanho Kim}
\email{}
\affiliation{Department of Astronomy and Space Science, Chungbuk National University, Cheongju, 28644, Republic of Korea}
\author[0000-0002-9709-5389]{Kaya Mori}
\email{}
\affiliation{Columbia Astrophysics Laboratory, 550 West 120th Street, New York, NY 10027, USA}
\author[0000-0001-6189-7665]{Samar Safi-Harb}
\email{}
\affiliation{Department of Physics and Astronomy, University of Manitoba, Winnipeg, MB R3T 2N2, Canada}


\begin{abstract}
We present a comprehensive X-ray study of the pulsar wind nebula (PWN) CTB 87, utilizing new NuSTAR observations and archival Chandra and XMM-Newton data. Our joint analysis reveals that the PWN's X-ray spectrum exhibits significant curvature, well described by a broken or cutoff power law. The broken power-law fit yields photon indices $\Gamma_{1}=1.89\pm 0.05$ and $\Gamma_{2}=2.73 \pm 0.10$ below and above a break energy of $E_{\rm b}=5.12 \pm 0.26$\,keV, respectively, while the cutoff power-law fit gives
$\Gamma=1.23 \pm 0.10$ and $E_{\rm c} = 5.29 \pm 0.67$\,keV. Spatially resolved spectroscopy confirms a previously observed radial softening of $\Gamma$, indicating significant synchrotron cooling as electrons move outward from the pulsar. We employed leptonic emission scenarios to model the broadband emission from CTB~87, focusing on the measured X-ray spectrum and the radial profiles. While an evolutionary one-zone model can explain the general features of the broadband spectral energy distribution (SED), its predictions do not fully reproduce the observed X-ray spectrum. Our multi-zone modeling demonstrates that the hard electron distribution required to explain the X-ray spectrum cannot account for the observed gamma-ray SED. This discrepancy suggests the presence of an additional electron population, possibly linked to the reverse shock interaction, and highlights CTB~87 as an evolved PWN whose broadband emission cannot be captured by these simple SED models. We discuss the potential origins for this second electron population and its implications for understanding the evolution of PWNe.
\end{abstract}

\bigskip 
\section{Introduction}
\label{sec1}
Pulsar wind nebulae (PWNe) are powered by relativistic particle winds emanating from rotation-powered pulsars, offering crucial insights into energetic phenomena within the Galaxy. At the termination shock \citep[e.g.,][]{kc84a}, formed by the interaction of the pulsar wind with the surrounding supernova remnant (SNR) or the interstellar medium (ISM), relativistic electrons and positrons are accelerated to extreme energies \citep[e.g.,][]{Cao2024}. As these particles travel through the PWN, they lose energy by emitting synchrotron and inverse-Compton (IC) radiation, which we observe across the electromagnetic spectrum from radio waves to gamma rays. This evolutionary process substantially modifies the electron energy distribution and consequently PWN properties with time. Notably, a distinct population of high-energy electrons can be created during a reverse-shock interaction phase, as a result of rapid cooling \citep[e.g.,][]{Slane2010}.

The broadband spectral energy distributions (SEDs) of PWNe serve as direct probes of the energy distribution of their emitting electrons. SED data from regions close to the pulsar, where particles are initially injected, reveal the properties of shock-accelerated electrons. Conversely, data from the outer regions provide crucial insights into how these particles flow and lose energy as they propagate through the nebula. Additionally, the magnetic-field structure, which can be measured via X-ray polarimetry \citep[e.g.,][]{Bucciantini2023,Romani2023}, offers important information on the transport of the most energetic electrons. Consequently, PWNe are invaluable cosmic laboratories for investigating particle acceleration and the magnetohydrodynamic (MHD) flows of relativistic particles \citep[see][for a review]{Gaensler2006}.

Radio and X-ray emissions from PWNe, originating from synchrotron processes, provide direct probes of the energetic electrons within them. Features in the X-ray spectrum, such as a break or a cutoff, provide clues to magnetic-field strength ($B$) and the maximum energy of these particles, while radio emission helps understand the lower-energy electron population. Gamma-ray emission arises from IC scattering of these electrons off ambient photon fields. Since the electrons responsible for X-rays also contribute to very-high-energy (VHE; $>100$\,GeV) gamma-ray emission, jointly modeling the X-ray and VHE SEDs is essential for a comprehensive understanding of the energetic electrons within PWNe and their evolution.

While broadband SED data are invaluable, other measurements offer further information. Radial profiles of X-ray spectral properties can aid in understanding the energy-loss processes and probe the $B$ structure and flow properties within PWNe \citep[e.g.,][]{Park2023}. Understanding temporal evolution, which also affects the particle distribution, is more challenging due to the limited number of samples with clear signatures of evolution. Nevertheless, the presence of relic electrons in some PWNe, such as the Vela-X PWN, may offer clues to their long-term evolution \citep[e.g.,][]{Slane2018}. Therefore, it is important to study PWNe which exhibit all these ingredients: broadband emission, a spatially varying X-ray spectrum, and evidence for relic electron populations.

\begin{figure}[t]
\begin{center}
\includegraphics[width=3.32 in]{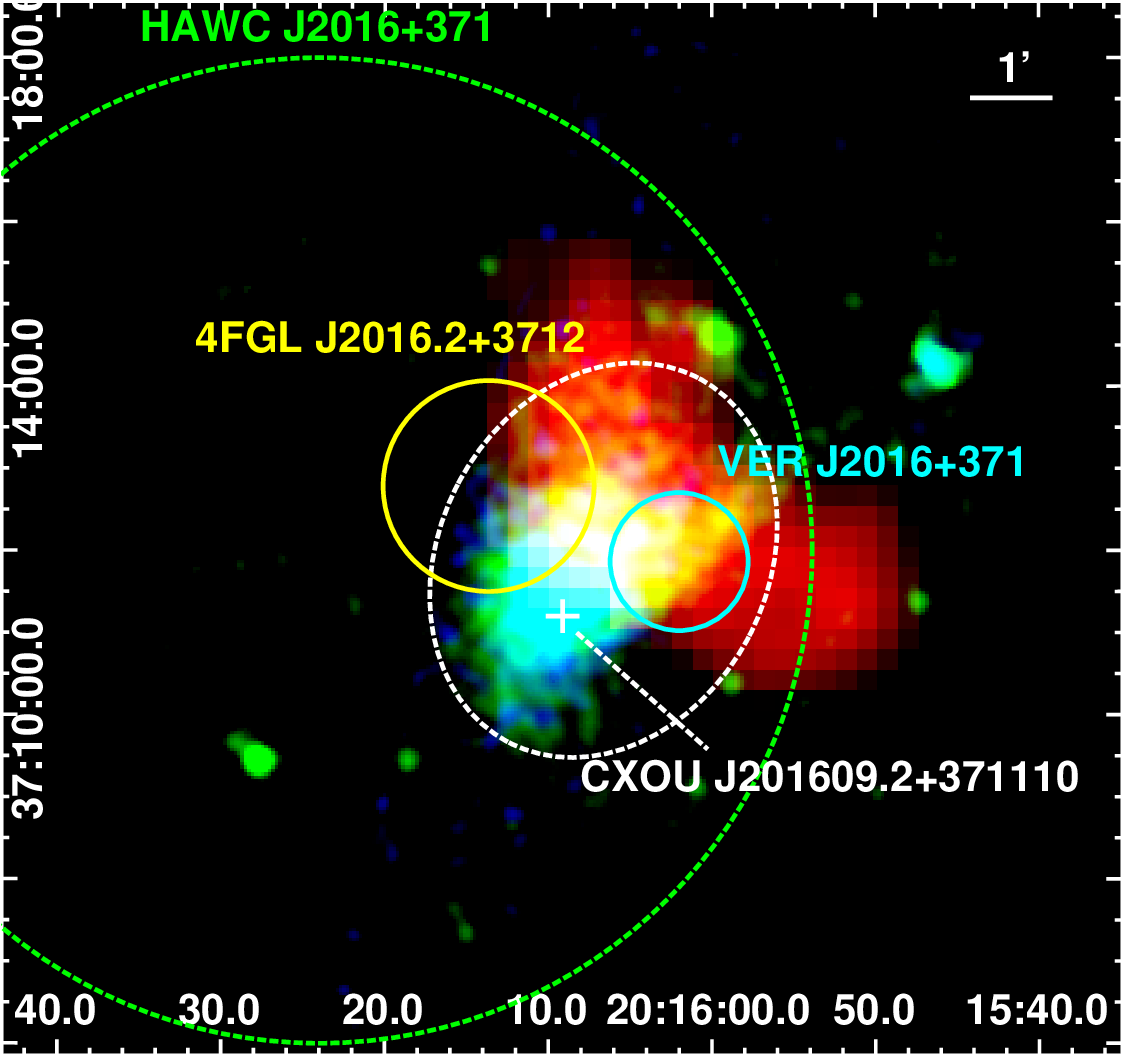}
\vspace{-7mm}
\end{center}
\caption{A composite image of CTB~87. VLA, XMM-Newton (2--7\,keV), and NuSTAR (3--20\,keV) images are depicted in red, green, and blue, respectively. Positions of its gamma-ray counterparts 4FGL J2016.2+3712 \citep[][]{Ballet2023}, VER~J2016+371 \citep[][]{Abeysekara2018}, and HAWC J2016+371 \citep[][]{Albert2021} are marked by yellow, cyan, and green circles (their radii are $1\sigma$ positional uncertainties), respectively. The white ellipse delineates the $2'\times 2.5'$ (called `PWN') region analyzed in Section~\ref{sec2_4_2}, while the Chandra counterpart of PSR J2016+3711 is indicated with a cross.}
\vspace{0.5cm}
\label{fig1}
\vspace{-0.8cm}
\end{figure}

The PWN CTB 87 (G74.9+1.2)\footnote{http://snrcat.physics.umanitoba.ca/SNRtable.php} is a compelling target for studying PWN physics, possessing the characteristics necessary for investigating its spatial and temporal evolution. It exhibits multiwavelength emission across the radio to VHE band \citep[Figure~\ref{fig1};][]{Kothes2020,Ballet2023,Aliu2014} and has well-constrained X-ray radial profiles \citep[][]{Guest2020}. The occurrence of a reverse shock interaction, inferred from its radio and X-ray properties \citep[][]{Matheson2013,Kothes2020}, suggests a population of relic electrons, offering a window into its past evolutionary history. The power source of this PWN was recently discovered at radio wavelengths by FAST and identified as PSR J2016+3711 (hereafter J2016) with a spin period of 50.81\,ms (19.68165\,Hz), a spin-down luminosity of $2.2\times10^{37}\,\mathrm{erg}\ \mathrm{s}^{-1}$, and a characteristic age of 11.2\,kyr \citep[][]{Liu2024}.

In the radio band, CTB 87 exhibits a kidney-shaped morphology \citep[][]{Kothes2020}, offset by $\sim100''$ from the pulsar (Figure~\ref{fig1}). The X-ray emission exhibits a compact $R\approx 5''$ inner nebula (Figure~\ref{fig2}) with spectrally hard emission ($\Gamma=1.2\pm0.5$) and an extended $R\approx 2.5'$ nebula \citep[][]{Matheson2013}. This extended emission has been well characterized by a power law with a photon index of $\Gamma=1.7$ or 1.9, as measured by Chandra \citep[][]{Matheson2013} and XMM-Newton \citep[][]{Guest2020}, respectively. Consistent with particle energy losses, this X-ray emission is observed to soften with increasing distance from the pulsar. In the gamma-ray band, 4FGL J2016.2+3712 \citep[][]{Ballet2023}, VER~J2016$+$371 \citep[][]{Aliu2014}, and HAWC~J2016+371 \citep[][]{Albert2021} are considered counterparts to CTB 87. The gamma-ray SED exhibits a complex shape spanning a wide energy range between $\sim$1\,GeV and $\sim$10\,TeV. 

Previous SED studies of CTB 87 have modeled its broadband emission using either a lepto-hadronic \citep[][]{Saha2016,Xin2025} or a purely leptonic scenario \citep[][]{Gong2025}. However, these studies were limited by their reliance on simple power-law fits to existing Chandra and XMM-Newton data. These $<$10\,keV data may have failed to capture high-energy features such as spectral curvature or cutoffs, which have been detected in other PWNe by NuSTAR \citep[e.g.,][]{An2019,Bamba2022}. Consequently, the electron distribution at the highest energies remains largely unconstrained. Given that CTB 87 is an evolved PWN that has likely undergone significant synchrotron cooling and reverse-shock compression, an accurate X-ray characterization and a refined SED model are crucial for understanding its evolution.

In this paper, we present a comprehensive X-ray analysis of CTB 87 using archival Chandra and XMM-Newton data, along with new NuSTAR observations. We measure the X-ray spectrum and the radial profiles of $\Gamma$ and surface brightness of the PWN (Section~\ref{sec2}). These X-ray data are then integrated with published radio and gamma-ray measurements to model the PWN emission, offering new insights into the PWN's evolution (Section~\ref{sec3}). We discuss our results in Section~\ref{sec4} and provide a summary in Section~\ref{sec5}.

\section{X-ray Data Analysis}
\label{sec2}

\subsection{Data reduction}
\label{sec2_1}
We analyzed X-ray data of CTB~87 obtained from archival Chandra \citep[][]{Garmire2003} and XMM-Newton \citep[][]{Jansen2001} observations, as well as new NuSTAR \citep[][]{Harrison2013} observations (Table~\ref{ta:ta1}). Two Chandra observations, acquired on 2001 July 8 for 18\,ks (Obs. ID 1037, ACIS-S) and 2010 January 16 for 69\,ks \citep[Obs. ID 11092, ACIS-I;][]{Matheson2013}, were reprocessed using the {\tt chandra\_repro} tool within CIAO 4.16.

The XMM-Newton data were acquired on 2014 December 14 \citep[Obs. ID 0744650202;][]{Guest2020} in full-frame mode for both the MOS and PN detectors. We processed these data with the {\tt emproc} and {\tt epproc} tools of XMM-SAS v22.1.0. Further removal of soft proton contamination was performed using {\tt espfilt}. The resulting XMM-Newton exposures were 98\,ks, 101\,ks, and 75\,ks for MOS1, MOS2, and PN, respectively.

NuSTAR observed CTB~87 on 2018 January 11 (Obs. ID 30460021002) and 2024 October 8 (Obs. ID 41001002002). In the earlier observation, the source was serendipitously detected at a $5'$ off-axis angle, whereas it was positioned on-axis in the 2024 observation. We processed these data using {\tt nupipeline} in HEASOFT v6.34 along with the optimized SAA filter.

\newcommand{\marka}{\tablenotemark{\tiny{\rm a}}}
\newcommand{\markb}{\tablenotemark{\tiny{\rm b}}}
\newcommand{\markc}{\tablenotemark{\tiny{\rm c}}}
\newcommand{\markd}{\tablenotemark{\tiny{\rm d}}}
\newcommand{\marke}{\tablenotemark{\tiny{\rm e}}}
\newcommand{\markf}{\tablenotemark{\tiny{\rm f}}}
\newcommand{\markg}{\tablenotemark{\tiny{\rm g}}}

\begin{table}[t]
\begin{center}
\caption{Data used in this work}
\label{ta:ta1}
\scriptsize{
\hspace{-15 mm}
\begin{tabular}{lcccc} \hline\hline
Instrument & Obs. ID(s) & Exposure \\ \hline
 &  & (ks) \\ \hline
Chandra  & 1037, 11092\marka & 18, 69 \\
XMM-Newton  & 0744650202 & 98, 101, 75\markb \\
NuSTAR  & 30460021002\markc, 41001002002 & 59, 63 \\\hline
\end{tabular}}
\end{center}
\vspace{-2.0 mm}
\footnotesize{
\marka{Data contained in \href{https://doi.org/10.25574/cdc.456}{Chandra Data Collection 456}}. \\ 
\markb{For MOS1, MOS2, and pn.}\\
\markc{CTB 87 was observed at a large off-axis angle of $5'$.}\\}
\end{table}

\subsection{Timing analysis}
\label{sec2_2}
To search for X-ray pulsations from the pulsar J2016, we analyzed data from the NuSTAR observations. We corrected the photon arrival times to the solar system barycenter using the Chandra-derived position (R.A. = $20^{\rm h}16^{\rm m}09\fs2$, Decl. = $+37^{\circ}11'10''$). Source events were extracted from a circular region with a $60''$ radius centered on this position. This extraction yielded 2200 and 3100 counts in the 3--20\,keV band from the 2018 and 2024 observations, respectively.

We performed a pulsation search around the predicted spin frequencies of $f=19.68346$\,Hz and 19.67751\,Hz which were extrapolated from the timing solution of \citet{Liu2024} for the 2018 and 2024 NuSTAR observation epochs. We searched a frequency interval of $\Delta f=\pm 10^{-4}$\,Hz (significantly larger than the frequency uncertainties), using the $H$-test \citep[][]{drs89} and maximizing the $H$ value by varying the extraction region and energy band. This analysis yielded no prominent peak in the $H$ values within the searched range, indicating no significant detection of X-ray pulsations. The maximum $H$ values were 13 (2018 data) and 15 (2024 data), corresponding to post-trial chance probabilities of $p\approx 0.5$ and $\approx 0.3$, respectively.

\begin{figure*}[t]
\begin{center}
\begin{tabular}{cccc}
\includegraphics[width=1.65 in]{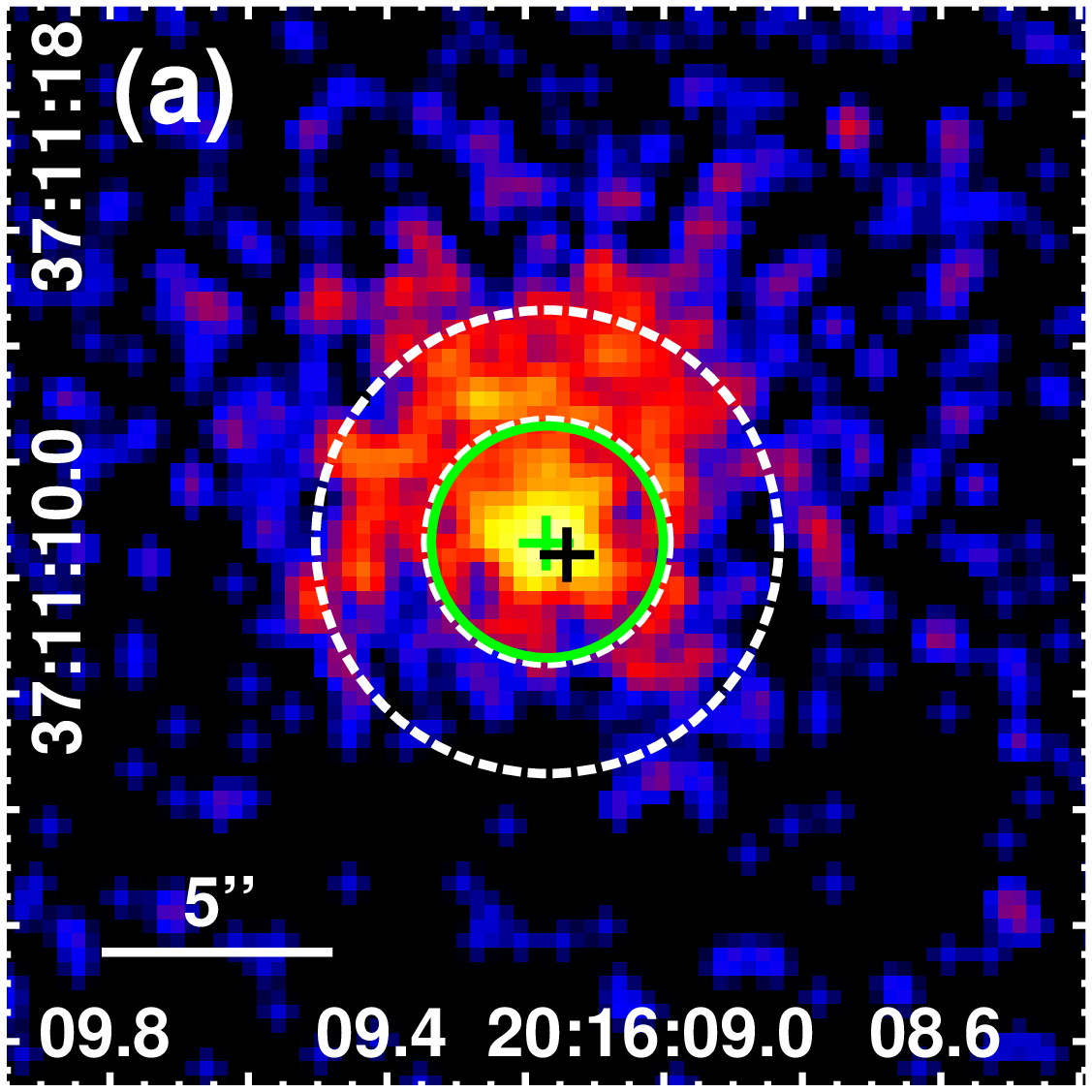} &
\includegraphics[width=1.65 in]{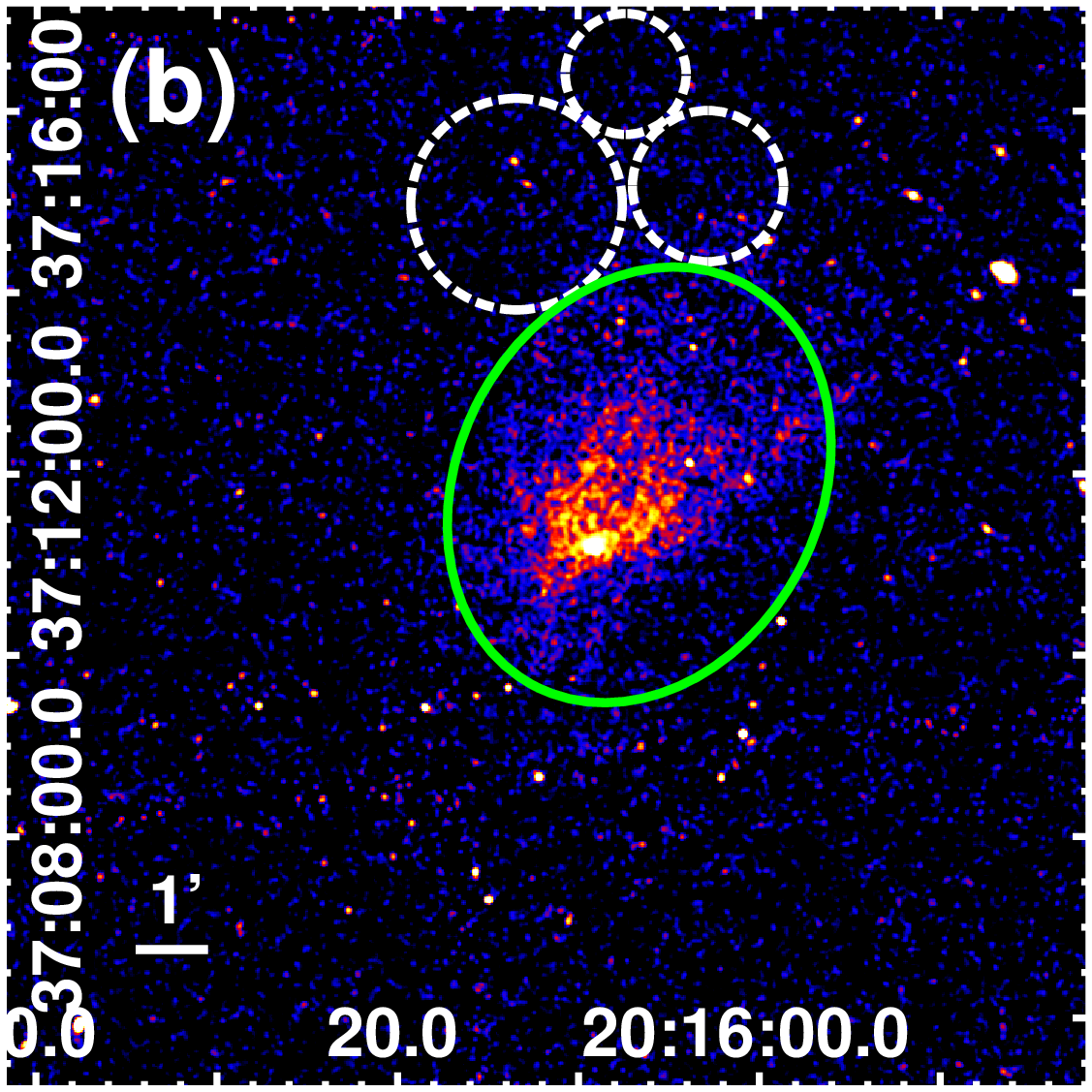} &
\includegraphics[width=1.65 in]{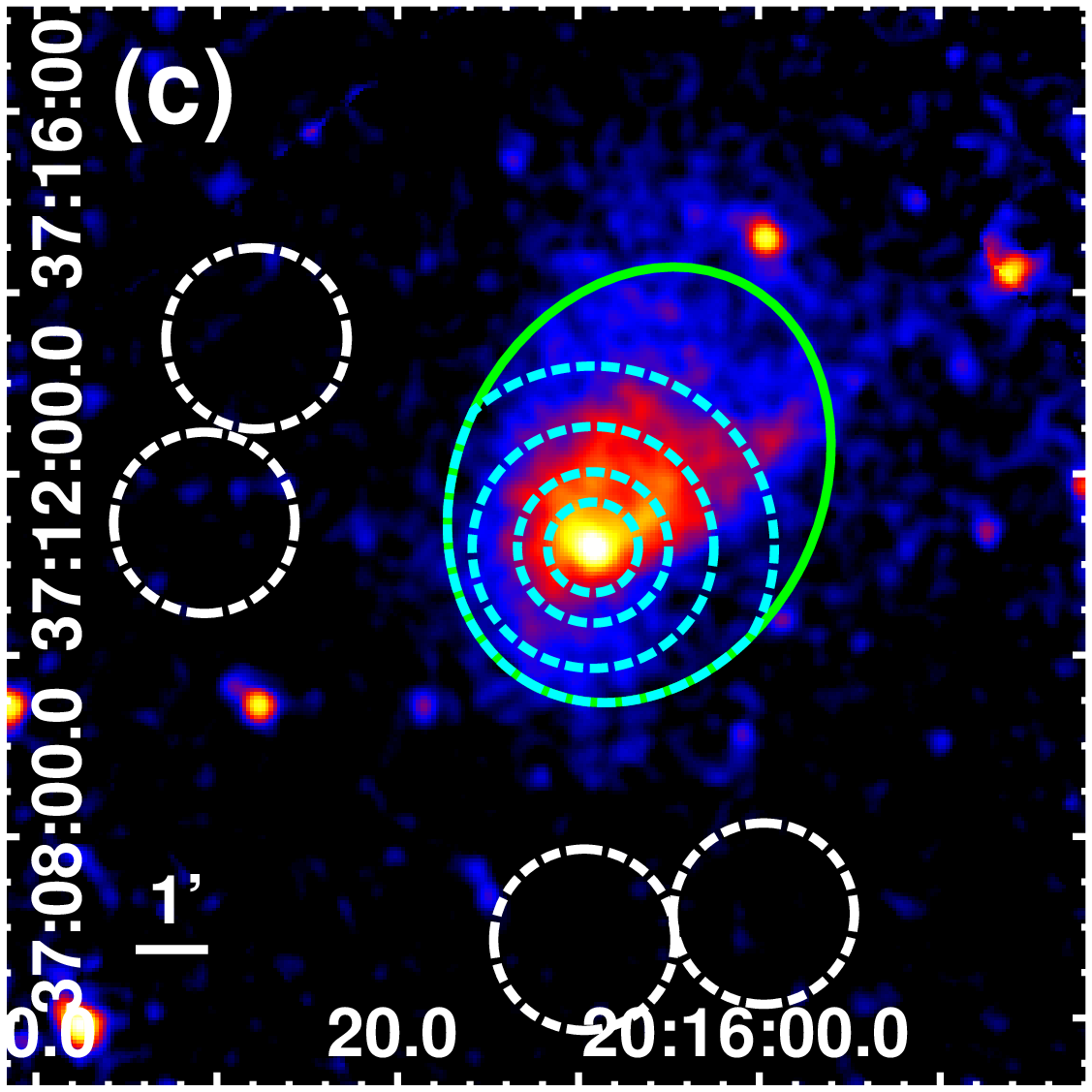} &
\includegraphics[width=1.65 in]{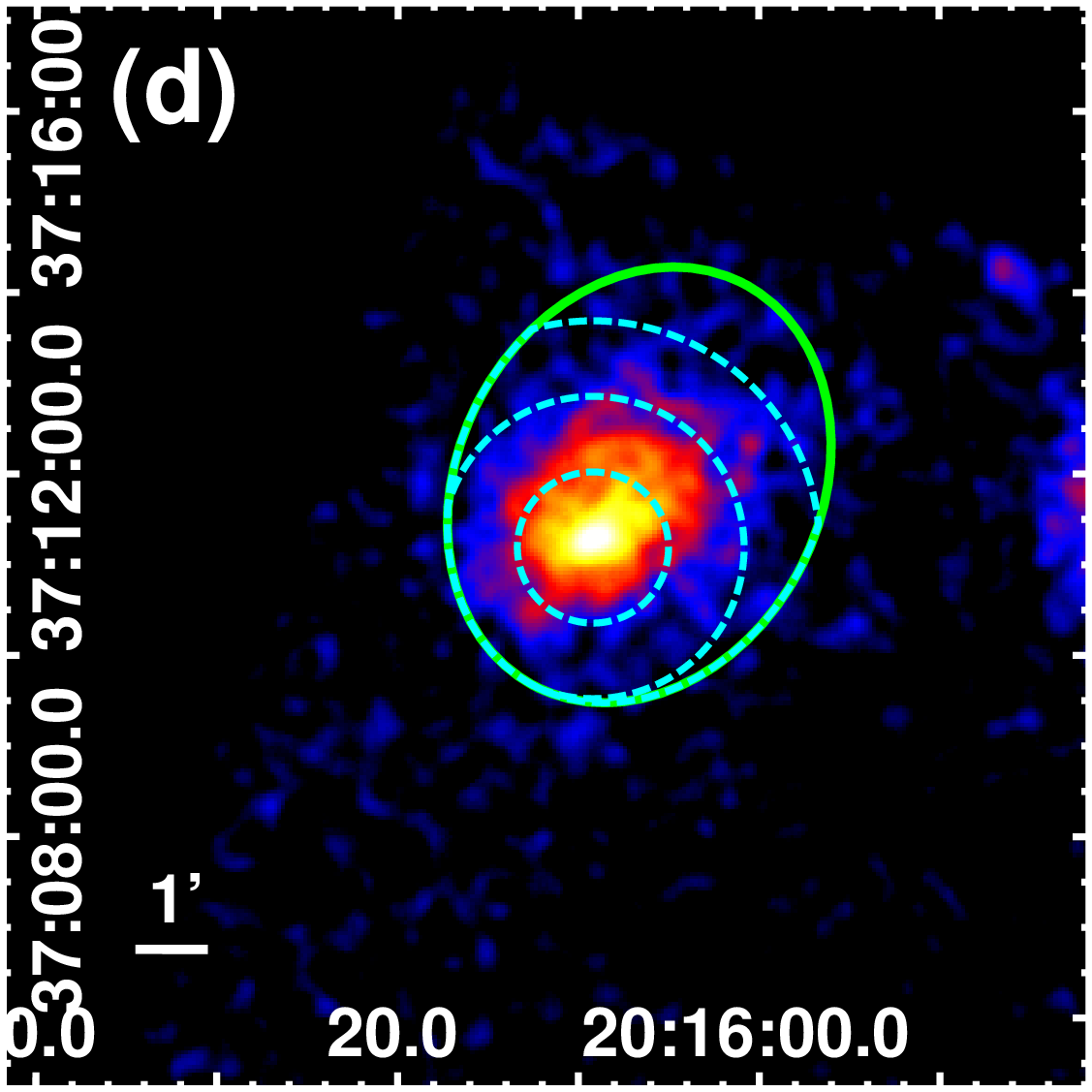} 
\vspace{-5mm}
\end{tabular}
\end{center}
\caption{(a) Chandra 1--7\,keV image of CTB~87, showing the small-scale inner structures. 
The green and black crosses mark the positions of the point source CXOU~J201609.2+371110 
and the radio pulsar J2016, respectively. (b), (c), and (d): PWN images on a larger scale generated from Chandra (1--7\,keV), XMM-Newton (1--7\,keV), and NuSTAR (3--20\,keV) observations, respectively. The green solid and white dashed regions denote the source and background extraction areas used for spectral analysis, respectively. The cyan dashed concentric annuli within the green regions in panels (c) and (d) show the regions used for the spatially resolved spectral analysis described in Section~\ref{sec2_4_3}.}
\vspace{0.5cm}
\label{fig2}
\vspace{-0.5cm}
\end{figure*}

\subsection{Image analysis}
\label{sec2_3}
We constructed X-ray images of CTB 87, displayed in Figure~\ref{fig2}. The high-resolution Chandra data reveal structures across various scales, previously identified by \citet{Matheson2013}. While these structures are significantly blurred in the XMM-Newton and NuSTAR data, the extended PWN of J2016 remains clearly visible. We investigated these images across different energy bands to search for changes in the PWN size that would indicate synchrotron cooling.

The Chandra and the NuSTAR data do not show a significant change in PWN size with energy, likely due to their relatively low photon counts. However, the high-quality XMM-Newton data are sufficient to distinguish between the PWN's morphology in low- and high-energy bands.

\begin{figure}
\begin{center}
\hspace{-2 mm}
\includegraphics[width=1.65 in]{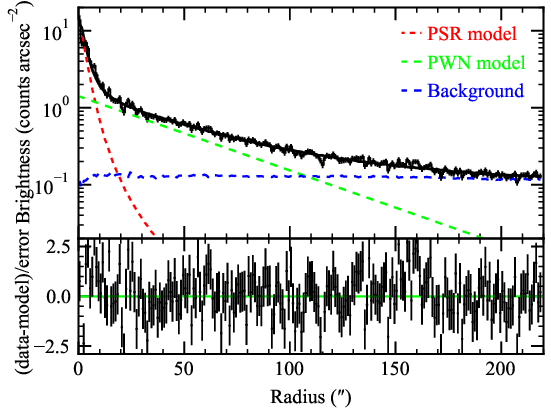}
\includegraphics[width=1.65 in]{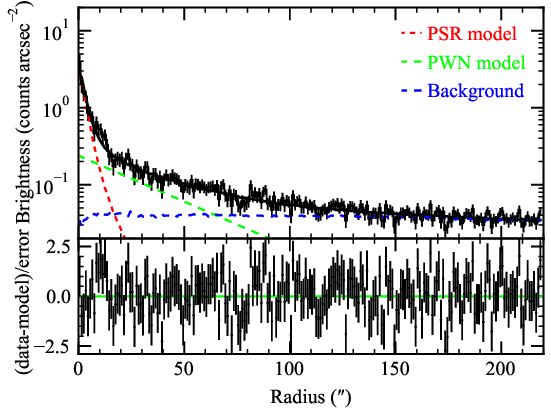}
\vspace{-5mm}
\end{center}
\caption{XMM-Newton radial surface brightness profiles of the PWN in the 1--5\,keV (left) and 5--10\,keV (right) bands. The best-fit model components are shown by the dashed curves: pulsar (red), PWN (green), and background (blue). The solid black curve represents the summed model.
}
\vspace{-0.4cm}
\label{fig3}
\end{figure}

To quantify the size change observed with XMM-Newton, we analyzed radial profiles. To best illustrate this effect, we selected the 1--5\,keV and 5--10\,keV bands and constructed corresponding radial profiles (Figure~\ref{fig3}). The figure clearly demonstrates that the PWN extends to a larger radius in the softer (1--5\,keV) band, providing evidence for synchrotron cooling within the source.

We then modeled these profiles to precisely characterize the energy-dependent size change. The pulsar component was represented by the point spread function (PSF), which was generated using the SAS tool \texttt{psfgen}. The PWN component was modeled by a simple exponential function: $A \mathrm{exp}(-r/l)$. We assumed a flat background, which was modeled using the exposure map. The amplitudes of these components and the exponential width ($l$) of the PWN were optimized to fit the observed radial profiles.

The fit results are presented in Figure~\ref{fig3}. The measured values of $l$ are $45''.0\pm1''.0$, and $36''.4\pm2''.3$ for the low- and high-energy bands, respectively. This difference is statistically significant with a confidence level of $\approx 3.4\sigma$, thereby demonstrating the expected shrinkage of the PWN size with increasing energy.

\subsection{Spectral analysis}
\label{sec2_4}
We conducted a comprehensive X-ray spectral analysis of the pulsar J2016 and its PWN using data from Chandra, XMM-Newton/MOS, and NuSTAR. We first measured the pulsar spectrum using the high-spatial resolution Chandra data (Section~\ref{sec2_4_1}). We then analyzed the Chandra, XMM-Newton and NuSTAR observations to determine the broadband X-ray spectrum of the PWN, accounting for the contamination from the pulsar (Section~\ref{sec2_4_2}). Finally, we performed a spatially-resolved spectral analysis of the PWN to measure radial profiles of $\Gamma$ and surface brightness of the PWN (Section~\ref{sec2_4_3}).

We extracted source spectra from regions shown in Figure~\ref{fig2} and generated the corresponding spectral response files for the point (pulsar) and extended (PWN) sources using the observatory-specific tools: {\tt specextract} for Chandra, {\tt arfgen} and {\tt rmfgen} for XMM-Newton, and {\tt nuproducts} for NuSTAR. Backgrounds for the Chandra and XMM-Newton data were extracted from source-free regions, whereas NuSTAR background was estimated using {\tt nuskybgd} simulations \citep[][]{Wik2014} due to stray-light contamination and the lack of a sufficient source-free region on the same detector chip.

Unless otherwise noted, we grouped the spectra to ensure at least 30 counts per bin and performed spectral fitting in XSPEC v12.14.0, employing the $\chi^2$ statistic. For joint analysis of these data, we introduced a cross-normalization constant to each dataset to account for potential cross-calibration issues \citep[e.g.,][]{Madsen2017}. We adopted the {\tt tbabs} model with the {\tt wilm} abundances \citep[][]{wam00} and the {\tt vern} cross section \citep[][]{vfky96} for Galactic absorption. Reported uncertainties correspond to the 1$\sigma$ confidence intervals.

\begin{table*}[t]
\vspace{-0.0in}
\begin{center}
\caption{Spectral analysis results}
\label{ta2}
\vspace{-0.05in}
\scriptsize{
\begin{tabular}{lccccccccc} \hline\hline
        Data       & {Instrument}\marka & Model  & Energy & $N_{\rm H}$  & $\Gamma$ & $E_{\rm b}$/$E_{\rm c}$\markb & $\Gamma_2$  &  $F_{\rm 2-10\,keV}$\markc & $\chi^2$/dof \\
           &   &     & (keV)   & ($10^{22}\rm \ cm^{-2}$) &  & (keV) &  & \\ \hline
PSR        & C   & PL & 0.5--10 & $2.38 \pm 1.19$   & $1.17\pm0.49$ &  $\cdots$  &  $\cdots$ & $0.07 \pm 0.01$ & 46/54 \\  \hline
PWN        & C  & PL & 0.5--10  & $2.22 \pm 0.18$  & $1.77\pm0.10$ &  $\cdots$  &  $\cdots$ & $2.06\pm0.07$  & 259/266 \\
PWN        & X  & PL & 0.5--10  & $2.24 \pm 0.07 {\pm 0.01}\markd$  & $2.02 \pm 0.04 {^{+0.01}_{-0.02}}$ &  $\cdots$  &  $\cdots$ & $1.87\pm0.03 {\pm0.01}$  & 301/272 \\
PWN        & N  & PL & 3--20   & 2.22\marke   & $2.74\pm0.10 {^{+0.10}_{-0.04}}$ &  $\cdots$  &  $\cdots$ & $2.54 \pm 0.20 {^{+0.08}_{-0.05}}$   & 271/286 \\ 
PWN        & X+C+N   & PL & 0.5--20 & $2.45 \pm 0.07 {\pm0.02}$  & $2.11 \pm 0.04 {\pm 0.02}$ &  $\cdots$  &  $\cdots$ & $1.87 \pm 0.03 {\pm 0.01}$ & 985/879 \\
PWN        & X+C+N   & BPL & 0.5--20 & $2.14 \pm 0.08 {\pm0.01}$  & $1.89\pm 0.05 {\pm0.01}$ & $5.12 \pm 0.26 {^{+0.04}_{-0.01}}$ & $2.73 \pm 0.10 {^{+0.08}_{-0.06}}$ & $1.76 \pm 0.03 {^{+0.02}_{-0.01}}$ & 915/877 \\
PWN        & X+C+N   & \tt{cutoffpl} & 0.5--20 & $1.85 \pm 0.09 {^{+0.02}_{-0.03}}$  & $1.23 \pm 0.10 {^{+0.03}_{-0.08}}$ &  $5.29 \pm 0.67 {^{+0.4}_{-0.5}}$  &  $\cdots$ & $1.75 \pm 0.03 {\pm 0.02}$ & 900/878 \\ \hline
\end{tabular}}
\end{center}
\vspace{-0.5 mm}
\footnotesize{
\marka{C: Chandra, X: XMM-Newton/MOS, N: NuSTAR.}\\
\markb{Break ($E_{\rm b}$) and cutoff ($E_{\rm c}$) energies for the BPL and {\tt cutoffpl} models, respectively.}\\
\markc{Absorption-corrected 2--10\,keV flux in units of $10^{-12}\rm \ erg\ s^{-1}\ cm^{-2}$.}\\
\markd{The second error is the systematic uncertainty associated with the pulsar model. See text.}\\
\marke{Fixed at the value obtained from the Chandra result for the PWN.}\\
}
\end{table*}

\subsubsection{Spectrum of the Pulsar J2016}
\label{sec2_4_1} 
We analyzed the Chandra observations, utilizing its high angular resolution to measure the X-ray spectrum of the pulsar J2016 (CXOU J201609.2+371110; Figure~\ref{fig2}a) while minimizing contamination from the surrounding PWN. We extracted the source spectra from a circular region with a radius of $2''$ centered on the pulsar and the background from an annular region with inner and outer radii of $2.1''$ and $4''$. Due to limited photon statistics, we grouped the spectrum to have at least five counts per bin and used the $l$ statistic \citep[][]{lstat}.

The spectrum of J2016 is well described by an absorbed power-law (PL) model, with best-fit parameters presented in Table~\ref{ta2}. While our $\Gamma$ and flux values are statistically consistent with the analysis of the same Chandra observations by \citet{Matheson2013}, our hydrogen column density ($N_{\rm H}$) value differs. We attribute this difference to the adoption of distinct abundance tables. By adopting the {\tt angr} abundances \citep[][]{angr89}, we obtained $N_{\rm H} = (1.6 \pm 0.8) \times 10^{22}\,\mathrm{cm}^{-2}$, which is consistent with their result for the PWN.

\subsubsection{Spatially-integrated spectrum of the PWN}
\label{sec2_4_2} 
For measurements of the PWN spectrum, we used a $2.5'\times 2'$ elliptical source region (Figure~\ref{fig2}). While weak, the pulsar emission ($\sim$4\% of the PWN; Table~\ref{ta2}) could potentially influence our results. The high-resolution Chandra data allowed us to effectively remove the pulsar by excising a circular region of $5''$, thereby minimizing its contamination. For the XMM-Newton and NuSTAR spectra, which include the pulsar emission, we modeled the data with two PL components: one for the PWN emission and another fixed to the pulsar parameters derived from the Chandra analysis (Section~\ref{sec2_4_1}). We note that the PWN region contains six contaminating point sources, identified in the Chandra data. These point sources are very faint, having summed counts of 85 in the  1--10\,keV band, which is much smaller than the pulsar counts of 200. Therefore, we neglect contamination from these faint point sources.

Results from analysis of individual Chandra, XMM-Newton, and NuSTAR observations are presented in Table~\ref{ta2}. For the XMM-Newton analysis, we iteratively adjusted the frozen pulsar parameters to use the common $N_{\rm H}$ value for the pulsar and PWN PL models. For the NuSTAR analysis, we held $N_{\rm H}$ fixed at $2.22 \times 10^{22}\,\mathrm{cm}^{-2}$ inferred from a PL fit of the Chandra data; varying this $N_{\rm H}$ value has a negligible impact on the NuSTAR results.

\begin{figure*}
\begin{center}
\includegraphics[width=2.3 in]{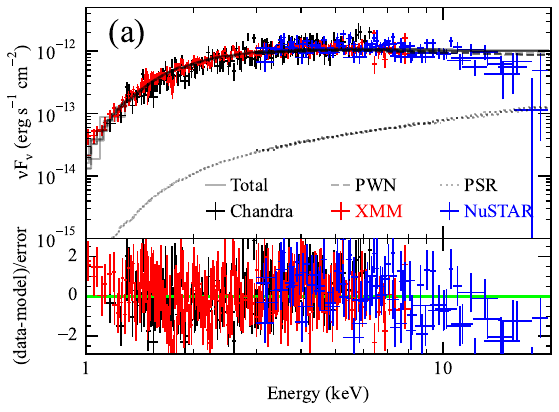}
\includegraphics[width=2.3 in]{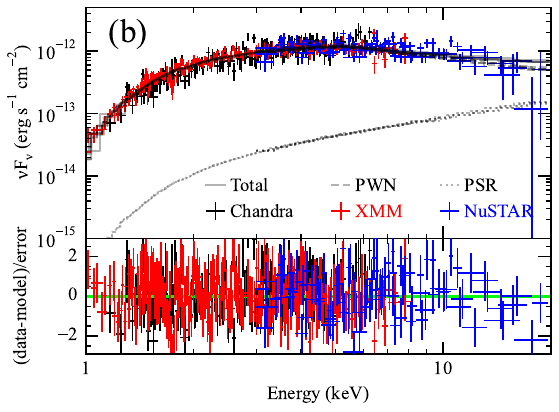}
\includegraphics[width=2.3 in]{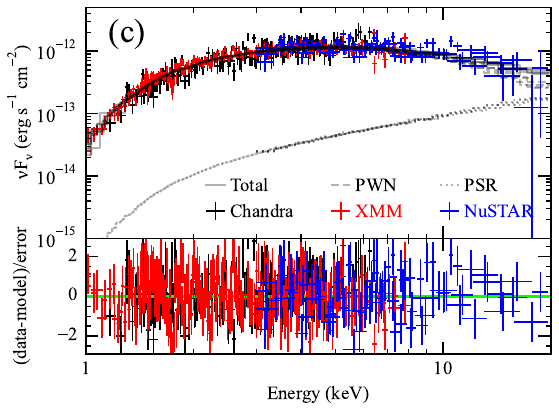}
\vspace{-5mm}
\end{center}
\caption{The 1--20\,keV spectrum of the PWN CTB 87 measured by Chandra (black points), XMM-Newton (red), and NuSTAR (blue). Panels (a), (b), and (c) display the best-fit results using the PL, BPL, and CPL models, respectively. Dotted, dashed, and solid curves represent the pulsar, PWN, and total emission components, respectively. The bottom panels show the residuals after subtracting the best-fit models.
}
\vspace{0.3cm}
\label{fig4}
\vspace{-0.2cm}
\end{figure*}

The distinct best-fit $\Gamma$ values obtained from individual fits suggest a potentially curved spectral shape. To investigate this further, we fit the XMM-Newton data with a broken power-law (BPL) model and found the best-fit photon indices of $\Gamma=1.89\pm0.05$ and $\Gamma_2=2.63\pm0.31$ below and above a break energy $E_{\rm b}=5.18\pm0.44$\,keV for $N_{\rm H}=(2.11\pm0.08)\times10^{22}\,\mathrm{cm}^{-2}$ ($\chi^2\text{/dof}=289/270$). This BPL model is favored over the simple PL model with a F-test probability of $\approx6\times10^{-3}$. The NuSTAR data also favored a BPL model (F-test probability of $\sim 8\times 10^{-5}$) with $\Gamma=2.31\pm0.06$, $\Gamma_2=4.75\pm1.17$, and $E_{\rm b}=11.4\pm0.92$\,keV ($\chi^{2}\textrm{/dof}=397/414$). The differences between the best-fit BPL parameter values from XMM-Newton and NuSTAR analyses possibly indicate that the spectrum exhibits a gradually curved shape rather than a single sharp break. 

Fitting each of the XMM-Newton and NuSTAR spectra with a {\tt cutoffpl} (CPL) model resulted in $N_{\rm H}=(1.78 \pm 0.12)\times10^{22}\,\mathrm{cm}^{-2}$, $\Gamma=1.15 \pm 0.21$, and a cutoff energy $E_{\rm c}=4.65\pm1.16$\,keV for XMM-Newton, and $\Gamma=1.34 \pm 0.29$ and $E_{\rm c}=5.79\pm1.51$\,keV with frozen $N_{\rm H}=1.78\times10^{22}\,\mathrm{cm}^{-2}$ for NuSTAR. For these datasets, the CPL model was favored over the PL model with F-test probabilities of $\approx 10^{-4}$ and $\approx 4\times10^{-5}$ for XMM-Newton and NuSTAR, respectively.

Since a joint analysis of the Chandra, XMM-Newton, and NuSTAR data would allow a more accurate characterization of the PWN's spectrum, we jointly analyzed the multi-instrument X-ray spectra utilizing a simple PL or curved PLs. A simple PL model resulted in a $\chi^2$ probability of $7\times10^{-3}$ (Table~\ref{ta2}) and leaves noticeable residuals; specifically, the model overpredicts the observed data at high energies (Figure~\ref{fig4}a). BPL and CPL models, displayed in Figure~\ref{fig4}b and c, provide acceptable fits with the $\chi^2$ probabilities of 0.18 and 0.30, respectively. The BPL and CPL models are significantly favored over the simple PL with F-test probabilities of $9\times 10^{-15}$ and $6\times 10^{-19}$, respectively, revealing significant spectral curvature in the X-ray spectrum of the PWN CTB~87. 

In order to assess the impact of the frozen PL parameters for the pulsar emission on the PWN results, we varied the pulsar parameters within their $1\sigma$ uncertainties. The curved PL models for the PWN spectra are always favored over the simple PL, regardless of the variations of the pulsar parameters. The changes in the PWN parameters induced by this variation are insignificant, smaller than their statistical uncertainties. We show these systematic uncertainties as the second errors in Table~\ref{ta2}.

We further verified that the weak pulsar emission did not affect the PWN results, by analyzing the data without excising the pulsar from the Chandra observation and without jointly modeling its emission in the XMM-Newton and NuSTAR analyses. We performed this analysis on individual and combined datasets. These data also require spectral curvature and the resulting parameter values for the PWN are fully consistent with those presented in Table~\ref{ta2}, verifying that our results are robust.

\subsubsection{Spatially-resolved spectra of the PWN}
\label{sec2_4_3}
To measure spatial variation of the X-ray emission from CTB 87, we performed a spatially-resolved analysis of the XMM-Newton and NuSTAR data. We defined the extraction regions using concentric annuli centered on the pulsar. To maintain consistency with our preceding spatially-integrated analysis (Section~\ref{sec2_4_2}), we selected only those portions of the annuli that overlap with the defined elliptical PWN region (Figure~\ref{fig2}c and d). For the XMM-Newton data, source spectra were extracted from five such overlapping regions defined by the following annuli radii: $0''\textrm{--}30''$, $30''\textrm{--}50''$, $50''\textrm{--}80''$, $80''\textrm{--}120''$, and $120''\textrm{--}200''$. For the NuSTAR analysis, larger widths for annular regions were adopted due to its lower spatial resolution and smaller photon counts: $0''\textrm{--}50''$, $50''\textrm{--}100''$, and $100''\textrm{--}150''$. The source emission outside $R=150''$ is not discernible from the background in the NuSTAR data (Figure~\ref{fig2}d). We note that there are a few contaminating point sources identified at $\gapp 150''$ in the XMM-Newton data; these were excised using $R=16''$ circles. However, we did not remove these sources from the NuSTAR data since they are undetected above 3\,keV.

\begin{figure}
\begin{center}
\includegraphics[width=3.25 in]{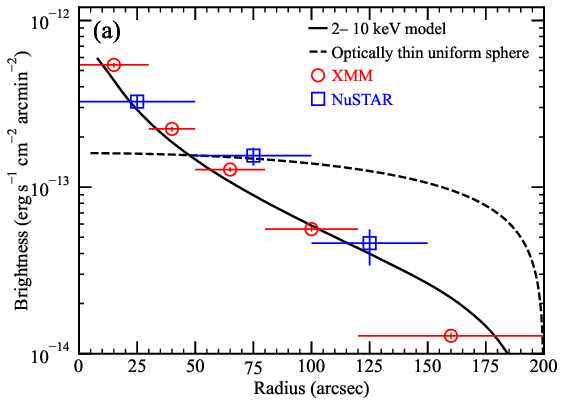} \\
\hspace{1.0mm}
\includegraphics[width=3.2 in]{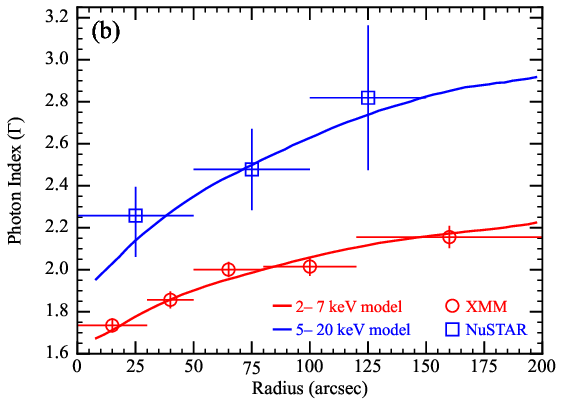}
\vspace{-5mm}
\end{center}
\caption{Radial profiles of surface brightness (a) and photon index (b) of CTB 87 measured by XMM-Newton (red) and NuSTAR (blue). In panel (a), the dashed curve represents the expected surface brightness profile (arbitrarily scaled) from one-zone models (i.e., optically thin, uniform sphere). The solid lines depict results from our multi-zone modeling, as described in Section \ref{sec3_3}, for the XMM-Newton (red curve) and NuSTAR (blue curve) data, respectively.}
\vspace{0.3cm}
\label{fig5}
\vspace{-0.2cm}
\end{figure}

For this analysis, we held $N_{\rm H}$ fixed at $2.14 \times 10^{22}~\mathrm{cm}^{-2}$, determined from the BPL model of the spatially-integrated data (Section~\ref{sec2_4_2}). As was done in Section~\ref{sec2_4_2}, we modeled the spectra with two PLs for the PWN and the pulsar. To assess the impact of the pulsar model on the PWN parameters, we varied the frozen pulsar parameters within their $1\sigma$ uncertainties. We found that the pulsar contamination was significant in the innermost regions, changing the PWN $\Gamma$ by $\pm 0.04$ and $^{+0.17}_{-0.09}$ for XMM-Newton and NuSTAR, respectively. The pulsar contamination had negligible impacts on the spectral results for the outer annuli. Uncertainties due to variation of the pulsar model were added to the statistical uncertainties in quadrature, and the best-fit PWN parameters are displayed in Figure~\ref{fig5}.

\section{Modeling of the PWN emission}
\label{sec3}
Previous broadband SED modeling of CTB 87 \citep[][]{Saha2016,Xin2025,Gong2025} relied on X-ray SEDs derived from simple PL fits to Chandra or XMM-Newton data. However, our current measurements, encompassing a broader X-ray band, strongly suggest a curved spectral shape. This finding has significant implications for the interpretation of high-energy emission from CTB 87, potentially necessitating a revision of previous model results.

To construct a broadband SED of CTB 87 (Figure~\ref{fig6}), we generated an X-ray SED based on the CPL model. We verified that using the BPL model does not significantly alter the overall SED shape. We then combined this X-ray SED with published radio \citep{Kothes2020}, Fermi-LAT \citep{Ballet2023}, and VERITAS \citep{Aliu2014,Abeysekara2018} data. The radio emission comprises two distinct components, a compact (kidney-shaped) structure representing the relic PWN and a larger diffuse component associated with the undisturbed PWN \citep{Kothes2020}; we model only the compact component in this work. \citet{Albert2021} reported a higher TeV flux measured by HAWC than the VERITAS results. However, these measurements remain statistically consistent due to their large reported uncertainties. For our modeling, we prioritized the VERITAS data for two reasons. First, VERITAS provides data points, whereas HAWC only provides a PL representation of the spectrum since it did not detect the source in individual energy bands. Second, VER~J2016+371 better overlaps spatially with CTB 87 (Figure~\ref{fig1}). 

We initially applied an evolutionary one-zone model (Section~\ref{sec3_1}) to determine whether evolutionary effects could account for the observed features of the source, namely, the high radio flux, the hard X-ray spectrum, and the broad gamma-ray SED with a sharp feature at $\sim0.1\textrm{--}1$\,TeV. However, given that our X-ray measurements, specifically the radial profiles of surface brightness and $\Gamma$ (Figure~\ref{fig5}b), indicate spatially varying PWN properties, we subsequently applied a multi-zone SED model to both the radial profiles and the broadband SED. Detailed descriptions of the one-zone and multi-zone models are provided in \citet{Gelfand2009} and \citet{Park2023b}, with brief overviews presented in Sections~\ref{sec3_1} and \ref{sec3_2}.

\subsection{Evolutionary one-zone modeling}
\label{sec3_1}
We adopt the evolutionary one-zone model developed by \citet{Gelfand2009}, which accounts for the evolution of PWN-SNR systems through pressure balance between the PWN and its surrounding media (the SNR and the ISM). The energy injection from the pulsar is assumed to vary with time following
\begin{equation}
\label{eq:Edot}
\dot E_{\rm SD}(t)=\dot E_{\rm SD}(0) \left ( 1 + \frac{t}{\tau_0} \right )^{-\frac{n+1}{n-1}},
\end{equation}
where $\tau_0$ is the spin-down timescale, $n$ is the braking index of the pulsar, and $t_{\rm age}$ is the PWN age. 

This injected energy exerts pressure on the surrounding medium, causing the PWN to initially expand. However, this expansion decelerates over time due to the adiabatic and radiative energy losses of the PWN electrons. Additionally, the PWN can be significantly compressed by a reverse shock from the SNR shell. This dynamic evolution modifies both the particle distribution and the PWN $B$, affecting the resulting SED in a complex way that depends on the evolutionary stage (i.e., age) of the PWN. A more complete description of the model is provided in \citet{Gelfand2009}.

This model has been applied to several PWNe \citep[e.g.,][]{Woo2023,Pope2024}, successfully reproducing their observed broadband SEDs and providing crucial insights into the PWN and SNR systems. Intriguingly, \citet{Pope2024} demonstrated that evolutionary effects can generate complex SED shapes. Specifically, a distinct population of electrons can form when a reverse shock compresses the PWN $B$, resulting in a sudden cooling of previously injected electrons. This mechanism may help accommodate the complex gamma-ray SED and high radio flux observed in CTB~87.

To account for both the radio and X-ray emission within a single injected population of electrons, we utilized a BPL distribution for the injection spectrum:
\begin{equation}
\label{eq:bpl}
\frac{dN_e}{dE_e} \propto \begin{cases} E_e^{-p_1}, & \textrm{for } E_{e,\rm min} \le E_e \le E_{e,\rm brk} \\
 E_e^{-p_2}, & \textrm{for } E_{e,\rm brk} < E_e \le E_{e,\rm max}. \\
\end{cases}
\end{equation}
Electrons are injected at each time step, and their energy distribution evolves over time according to the dynamic interaction between the PWN and the surrounding media \citep[e.g.,][]{Reynolds1984}. 
After this evolution, we compute the synchrotron and IC emissions (CMB seed photons) from the present-day electron population and compare the results with the observational data.
We do not include a Galactic IR seed-photon field (e.g., $T_{\rm IR}\approx20$\,K and $U_{\rm IR}\approx0.5\,\mathrm{eV}\,\mathrm{cm}^{-3}$; Table~\ref{ta4}) in this model because we found that it does not improve the fit and its IC contribution is degenerate with the fitted distance and the $B$ energy fraction $\eta_B$ (Table~\ref{ta3}).
We fit the broadband SED using a Markov-chain Monte Carlo (MCMC) technique with  a Gaussian prior of $6.1\pm0.9$\,kpc on the distance, based on the HI absorption measurement of \citet{Kothes2003}.
All twelve parameters in the upper section of Table~\ref{ta3}  were fit simultaneously, with the pulsar spin period and period derivative  held fixed at their measured values \citep[][]{Liu2024}. 
The four quantities in the lower section of Table~\ref{ta3} are derived from the best-fit model.

\begin{figure}
\begin{center}
\hspace{-8 mm}
\includegraphics[width=3.4 in]{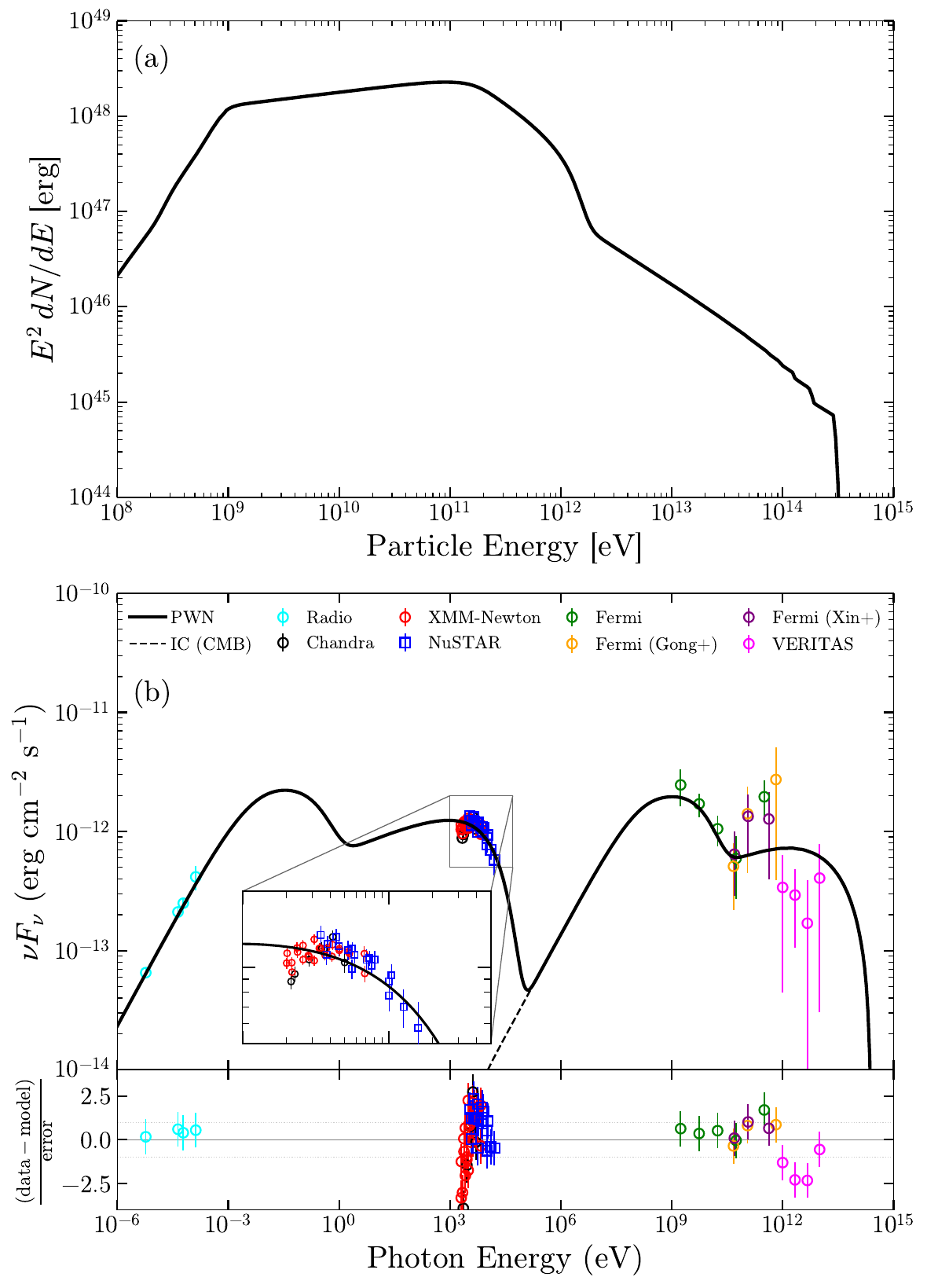}\\
\hspace{-8mm}
\vspace{-8mm}
\end{center}
\caption{One-zone modeling results. (a) Electron distribution used in the modeling. (b) Broadband SED of CTB~87 in the radio to TeV band. The data points include radio (cyan), and VERITAS (magenta) taken from literature \citep[][respectively]{Kothes2020, Abeysekara2018}, and our X-ray measurements (Section~\ref{sec2_4_2}). We also show Fermi-LAT analyses of \citet{Ballet2023}, \citet{Xin2025}, and \citet{Gong2025} as the green, purple, and orange circle, respectively. The solid black line is the SED predicted by the evolutionary one-zone model (Section~\ref{sec3_1}). The dashed curve shows the IC (CMB) contribution. The lower panel displays the fit residuals. 
}
\vspace{0.3cm}
\label{fig6}
\vspace{-0.2cm}
\end{figure}

The best-fit parameters are presented in Table~\ref{ta3}, and the resulting SED model is shown in Figure~\ref{fig6}. The model is broadly consistent with the observed SED,  given the measurement uncertainties, and the model-inferred values are typical for PWN-SNR systems. In particular, the supernova explosion energy, ejecta mass, and ISM density are consistent with ranges observed in other systems \citep[e.g.,][]{Pope2024,Gagnon2026}. Furthermore, the model-inferred distance and PWN age agree with independent estimates derived from radio and X-ray data \citep[e.g.,][]{Kothes2003,Matheson2013}.
Specifically, the best-fit  distance of 4.9\,kpc lies within just $\sim 1.3\sigma$ of the \citet{Kothes2003} HI distance, and this offset is driven by covariances of $d$ with $E_{\rm SN}$, $\eta_B$, $E_{e,\rm max}$, and the braking index.

\begin{table}[t]
\vspace{-0.0in}
\begin{center}
\caption{Parameters for the SED model in Figure~\ref{fig6}}
\label{ta3}
\vspace{-0.05in}
\scriptsize{
\begin{tabular}{lc} \hline\hline
Parameter & Value \\ \hline
\multicolumn{2}{l}{Fitted Parameters}\marka \\ \hline
SN explosion energy ($E_{\rm SN}$)        & $2.7\times 10^{50}$ erg   \\
Ejecta mass ($M_{\rm ej}$)                & 4.2 $M_\odot$     \\
ISM density ($n_{\rm ism}$)               & 5.6 $\rm cm^{-3}$  \\
Pulsar braking index ($n$)                & 3.39  \\
Spin-down time scale ($\tau_0$)           & 1430 yr   \\
$B$ energy fraction ($\eta_B$)            & $1.6\times 10^{-4}$         \\
Minimum electron energy ($E_{e,\rm min}$) & 0.950 GeV \\
Maximum electron energy ($E_{e,\rm max}$) & 325 TeV  \\
Break energy of particle spectrum ($E_{e,\rm brk}$) & 220 GeV  \\
Particle index below $E_{e,\rm brk}$ ($p_1$)        & 1.86   \\
Particle index above $E_{e,\rm brk}$ ($p_2$)        & 2.69   \\
Distance ($d$)\markb                       & 4.9 kpc   \\ \hline
\multicolumn{2}{l}{Derived Quantities} \\ \hline
Present $B$ ($B_{\rm present}$)            & 3.9 $\mu$G  \\
Present PWN radius ($R_{\rm PWN}$)         & 3.1 pc  \\
Present SNR radius ($R_{\rm SNR}$)         & 7.2 pc  \\
Age ($t_{\rm age}$)                        & 7.9 kyr  \\
\hline
\end{tabular}}
\end{center}
\vspace{-0.5 mm}
\footnotesize{
\marka{All twelve parameters were fit simultaneously. Additional input parameters are the pulsar spin period and period derivative, which were fixed at the  values measured by \citet{Liu2024}. Derived quantities are computed from the best-fit model.}\\
\markb{Fit with a Gaussian prior of $6.1\pm0.9$\,kpc \citep[HI absorption distance;][]{Kothes2003}; see Section~\ref{sec3_1}.}\\
}
\end{table}

While our model utilized a broad electron distribution similar to previous stationary one-zone models \citep[][]{Xin2025,Gong2025}, the detailed shape of the predicted SED differs. 
In our evolutionary model, the reverse shock reached the PWN at $\sim$1.6\,kyr and began compressing it at $\sim$2.0\,kyr, with the compression peaking at $\sim$4.7\,kyr, when the PWN $B$ reached $\sim$50\,$\mu$G, an order of magnitude above its present value (Table~\ref{ta3}).
This cooling creates a low-energy electron population that accounts for the radio and 1--10\,GeV flux. 
However, the predicted X-ray spectrum is broader and flatter than the observed one, not fully reproducing the measured spectral curvature (Section~\ref{sec2_4_2}), and the one-zone model also underpredicts the gamma-ray flux in the 0.1--1\,TeV band.

The one-zone model assumes that, at a given time, the electron population(s) responsible for the radio, X-ray, and gamma-rays move through regions with the same magnetic field strength.
In reality, electron population(s) could be flowing through regions with different magnetic field strengths. Given the complexity in the multi-band emission morphology and the broadband SED, these emissions may originate from multiple particle populations. 
Furthermore, the one-zone approach does not account for the spatially varying X-ray properties (Figure~\ref{fig5}). Therefore, we utilized a multi-zone model to provide a more comprehensive representation of the X-ray observations.

\subsection{Description of the multi-zone model}
\label{sec3_2}
We model the emission from the PWN using a multi-zone approach that accounts for the spatial evolution of electrons and their radiative output. This model assumes that electrons are injected at the termination shock with a prescribed energy distribution $dN_e/dE_e dt$ (e.g., a PL or Maxwellian). As these electrons propagate outward through advection and diffusion, their energies evolve under the influence of adiabatic and radiative cooling processes. This scenario has been applied to several PWNe and successfully explained the observed radial profiles and broadband SEDs \citep[e.g.,][]{Park2023, Kim2024}.

The total power injected into the PWN is determined by the pulsar's $\dot E_{\rm SD}$, which evolves over time according to Eq.~(\ref{eq:Edot}). For the spin-down timescale, we adopted $\tau_0=2\tau_c/(n-1) - t_{\rm age}$ with $n=3$. This energy is partitioned between electrons, the magnetic field, and radiation with fractions $\eta_e$, $\eta_B$, and $\eta_\gamma$ ($\eta_e + \eta_B + \eta_\gamma = 1$), respectively. The injected electron energy varies with time such that:
\begin{equation}
\label{eq:eevol}
\int_{E_{e,\rm min}}^{E_{e,\rm max}} E_e \frac{dN_e}{dE_e dt} dE_e = \eta_e \dot E_{\rm SD}(t).
\end{equation}
However, the PWN $B$ is assumed to be temporally constant, satisfying the energy condition:
\begin{equation}
\label{eq:totB}
\int_{R_{\rm TS}}^{R_{\rm PWN}} \frac{B(r)^2}{8\pi} 4\pi r^2 dr = \int_{0}^{t_{\rm age}} \eta_B \dot E_{\rm SD}(t),
\end{equation}
where $R_{\rm TS}$ and $R_{\rm PWN}$ are the radii of the termination shock and the synchrotron PWN, respectively. For $R_{\rm PWN}$, we adopt the X-ray size rather than the radio size because we model primarily the X-ray and VHE emissions. Radio emission can arise from a population of electrons that is distinct from those responsible for the X-ray and TeV emission (see Section~\ref{sec3_3}).

To match the observed spatially varying X-ray properties (Figure~\ref{fig5}), we assume that the bulk flow speed ($V_{\rm flow}$) of the electrons and the PWN's $B$ follow PL functions within $R_{\rm PWN}$ and remain constant beyond this radius:
\begin{equation}
\label{eq:vflow}
V_{\rm flow}(r)= \begin{cases}
 V_0 \left( \frac{r}{R_{\rm TS}} \right )^{\alpha_V}, & \textrm{for } R_{\rm TS} \le r \le R_{\rm PWN} \\
 V_{\rm ext}, & \textrm{for } R_{\rm PWN} < r \le R_{\rm max} \\
\end{cases}
\end{equation}
and
\begin{equation}
\label{eq:bfield}
B(r)= \begin{cases}
 B_0 \left( \frac{r}{R_{\rm TS}} \right )^{\alpha_B}, & \textrm{for } R_{\rm TS} \le r \le R_{\rm PWN} \\
 B_{\rm ext}, & \textrm{for } R_{\rm PWN} < r \le R_{\rm max}  \\
\end{cases}
\end{equation}
where $R_{\rm max}$ defines the boundary of our computational volume. We also incorporate electron diffusion with the following coefficient:
\begin{equation}
\label{eq:diff}
D(r,E_e)=D_0 \left( \frac{B(r)}{100\mu\rm G} \right )^{-1} \left( \frac{E_e}{511 \rm\ TeV} \right )^{1/3}.
\end{equation}

These properties are used to compute particle transport and energy losses, including synchrotron and adiabatic processes. For IC losses, we consider the cosmic microwave background (CMB) and a Galactic infrared (IR) field with a temperature $T_{\rm IR}$ and energy density $U_{\rm IR}$ as the seed photons.

The evolution of electrons and their emission are computed iteratively. At each time step ($dt$), electrons are injected at the termination shock and transported outward via advection and diffusion. The radial advective motion is analytically determined using Eq.~(\ref{eq:vflow}), while diffusion is modeled through simulations; electrons are randomly moved following a Gaussian distribution with $\sigma = \sqrt{2 D dt}$ in each of the $x$, $y$, and $z$ directions. Simultaneously, these electrons lose energy through adiabatic expansion (governed by Eq.~\ref{eq:vflow}) and radiative processes (synchrotron and IC). This procedure is repeated over $t_{\rm age}$, yielding the electron distribution at every position within our computational volume. Finally, we calculate their synchrotron and IC emissions, project them onto the observer's tangent plane, and then integrate the projected emission within appropriate radii for comparison with the observed radial profiles and broadband SED.

We should note that our multi-zone model has some limitations. While it accounts for the spatial variation of the PWN properties, it assumes a time-independent $B$ and $V_{\rm flow}$, and spherical symmetry. The assumption of a time-independent $B$ and $V_{\rm flow}$ is reasonable for the observed X-ray and $>$TeV emissions, as these arise primarily from the most recently injected, highest-energy electrons when $B$ and $V_{\rm flow}$ are not expected to change substantially. However, the spherically symmetric model does not fully capture the complex spatial variations of $B$ and $V_{\rm flow}$ within the asymmetric source. Therefore, the parameter values inferred from this model should be regarded as representing the angle-average properties of the PWN. Furthermore, strong covariance between the model parameters precludes precise determinations of their values; a previous study found that these values could vary by a factor of several due to this covariance \citep[e.g.,][]{Park2023b}. Consequently, the reported values represent one possible realization among many possibilities.

\subsection{Multi-zone modeling}
\label{sec3_3}

\begin{figure}
\begin{center}
\hspace{-3.5 mm}
\includegraphics[width=3.45 in]{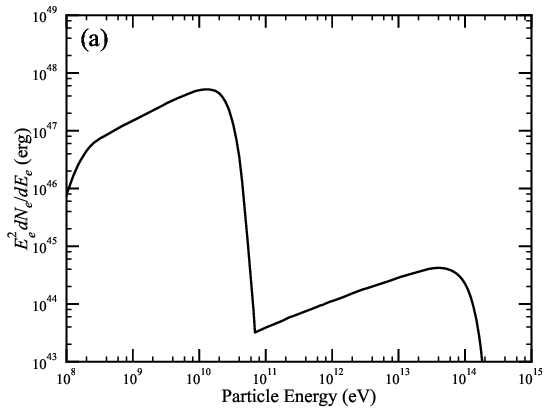}\\
\hspace{-4 mm}
\includegraphics[width=3.4 in]{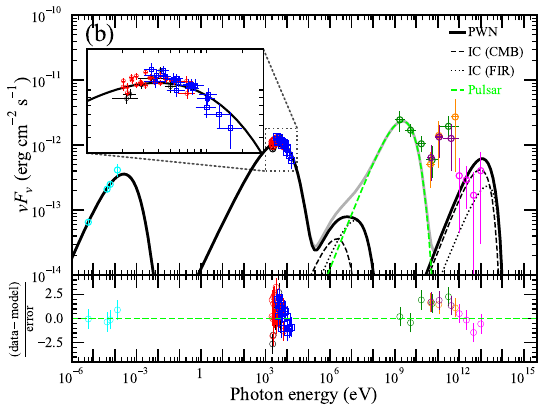}
\vspace{-6mm}
\end{center}
\caption{Multi-zone modeling results. (a) Spatially integrated electron distribution used in the modeling. (b) The resulting model SED (solid black line), with fit residuals displayed below. The green curve represents the potential gamma-ray emission from J2016, modeled using average spectral parameters of pulsars in the 4FGL catalog and normalized to the observed flux. Individual IC emission components from CMB (dashed) and far-IR (dotted) seed photons are also plotted.
}
\vspace{0.3cm}
\label{fig7}
\vspace{-0.2cm}
\end{figure}

Because the model includes many covarying parameters, constraining several of them in advance is crucial. We therefore estimated the initial parameter values for our multi-zone model using a one-zone approach. We adopted 6.1\,kpc for the distance to CTB 87 \citep{Kothes2003}. Based on this distance, the observed angular sizes of the X-ray torus and the PWN correspond to $R_{\rm TS}\approx 0.1$\,pc and $R_{\rm PWN}\approx5.9$\,pc, respectively. The size of the $\gapp$TeV emission region is typically dominated by particle diffusion ($\sim \sqrt{Dt_{\rm age}}$). Consequently, the point-like nature of the TeV counterpart VER J2016$+$371 suggests a small $D_0$ and a young age for the PWN.

The PWN age ($t_{\rm age}$) is highly uncertain, with previous studies suggesting a range of 5--28\,kyr \citep[][]{Matheson2013} or an upper limit of $<19$\,kyr \citep[][]{Guest2020}. The measured X-ray-to-TeV flux ratio of $\sim2$ suggests a potentially younger age of $\sim 3\textrm{--}6$\,kyr, based on established correlations inferred from SED modeling \citep{Zhu2018}. Consistent with these constraints, we adopt an age $t_{\rm age}=5$\,kyr for this work.

The observed emission at $\sim10$\,TeV, arising from IC scattering of CMB and IR seed photons, implies a maximum electron energy of $E_{e,\rm max}>100$\,TeV. The NuSTAR data at $\sim 20$\,keV then help estimate $B$ in the inner PWN to be $\sim 30\mu$G. The X-ray-to-TeV flux ratio of approximately 2 suggests an average $B$ of $\sim 5\mu$G. We estimate $\eta_\gamma < 0.01$ since the measured Fermi-LAT flux is $<1$\% of $\dot E_{\rm SD}$.

\begin{table}[t]
\vspace{-0.0in}
\begin{center}
\caption{Parameters for the SED model in Figure~\ref{fig7}}
\label{ta4}
\vspace{-0.05in}
\scriptsize{
\begin{tabular}{lc} \hline\hline Parameter & Value \\ \hline
\multicolumn{2}{l}{Fixed Parameters} \\ \hline
PWN age ($t_{\rm age}$) & 5 kyr \\
PWN radius ($R_{\rm PWN}$) & 5.9 pc \\
Termination-shock radius ($R_{\rm TS}$) & 0.09 pc \\
Maximum computation radius ($R_{\rm max}$) & 100 pc \\
Distance ($d$) & 6.1 kpc \\ \hline \multicolumn{2}{l}{Adjusted Parameters} \\ \hline 
Particle index ($p_1$) & 1.4 \\
Minimum electron energy ($E_{e,\rm min1}$) & 0.5 GeV \\
Maximum electron energy ($E_{e,\rm max1}$) & 8 GeV \\
Minimum electron energy ($E_{e,\rm min2}$) & 0.5 GeV \\
Maximum electron energy ($E_{e,\rm max2}$) & 180 TeV \\
Magnetic field ($B_0$) & 28.8 $\mu$G \\
Magnetic-field index ($\alpha_B$) & $-0.2$ \\
External magnetic field ($B_{\rm ext}$) & 2 $\mu$G \\
Flow speed ($V_0$) & 0.063 $c$ \\
Flow-speed index ($\alpha_V$) & $-0.8$ \\
External flow speed ($V_{\rm ext}$) & 0 $\mathrm{km}\,\mathrm{s}^{-1}$ \\
Diffusion coefficient ($D_0$) & $10^{27.38}\ \mathrm{cm}^2\,\mathrm{s}^{-1}$ \\
Particle energy fraction ($\eta_e$) & 0.97 \\
$B$ energy fraction ($\eta_B$) & 0.027 \\
CMB temperature ($T_{\rm CMB}$) & 2.7 K \\
CMB energy density ($U_{\rm CMB}$) & 0.26 $\mathrm{eV}\,\mathrm{cm}^{-3}$ \\
IR temperature ($T_{\rm IR}$) & 20 K \\
IR energy density ($U_{\rm IR}$) & 0.5 $\mathrm{eV}\,\mathrm{cm}^{-3}$ \\
\hline
\end{tabular}}
\end{center}
\vspace{-0.5 mm}
\end{table}

We subsequently refined these initial parameter values to reproduce the radial profiles of X-ray surface brightness and $\Gamma$ (Figure~\ref{fig5}) and  the broadband SED with our multi-zone model (Figure~\ref{fig7}). The resulting parameter values, optimized through visual inspection, are presented in Table~\ref{ta4}. While these values are comparable to those inferred for other PWNe \citep[e.g.,][]{Kim2024,ParkM2025}, the diffusion coefficient is substantially smaller than the average ISM value \citep[e.g.,][]{Yuan2017}. This is consistent with the slower diffusion near PWNe suggested for some sources associated with TeV halos \citep[e.g.,][]{Fang2018,Li2025}.

Notably, the hard PL electron distribution required to match the X-ray SED and the radial $\Gamma$ profile (Figure~\ref{fig5}b) did not produce sufficient radio flux. Consequently, we incorporated an additional PL distribution to account for the observed radio emission. The injected electron distribution is thus given by 
\begin{equation}
\label{eq:2PL}
\frac{dN_e}{dE_e dt} = \begin{cases}
 N_{e,1} E_{e}^{-p_1}, & \textrm{for } E_{e,\rm min1} \le E_e \le E_{e,\rm max1} \\
 N_{e,2} E_e^{-p_1}, & \textrm{for } E_{e,\rm min2} \le E_e \le E_{e,\rm max2}, \\
\end{cases}
\end{equation}
where the synchrotron emission from the first and second populations accounts for the observed radio and X-ray emission, respectively.

Our model successfully accounts for the radio, X-ray, and VERITAS data by the two electron populations (Eq.~\ref{eq:2PL}). However, the Fermi-LAT measurements remain unexplained (Figure~\ref{fig7}). While the low-energy $\lapp10$\,GeV emission could be attributed to an additional emission component (green in Figure~\ref{fig7}) such as gamma-ray emission from the pulsar \citep[Section~\ref{sec4_3}; see also][for alternative explanations]{Saha2016,Xin2025}, the high $\sim$TeV flux (i.e., bump at $\lapp \rm TeV$ energies in Figure~\ref{fig7}) measured by Fermi LAT presents a challenge.

Previous studies \citep{Gong2025,Xin2025} and our evolutionary one-zone model (Section~\ref{sec3_1}) addressed this TeV excess (i.e., TeV bump) by adopting broad electron energy distributions. However, these broad distributions are inconsistent with the hard $\Gamma$ \citep[Section~\ref{sec2_4_2}; see also][]{Matheson2013} and the curved X-ray spectrum revealed by our refined X-ray analysis; the curved spectrum requires a much harder and narrower electron distribution at high energies, which makes the $\sim$TeV excess even more prominent as shown in Figure~\ref{fig7}.

A potential solution to the TeV excess is to lower $E_{e,\rm max}$ to shift the IC SED peak ($\propto E_{e,\rm max}^2$) to approximately 1\,TeV. However, since the synchrotron emission frequency is proportional to $B E_e^2$, a lower $E_{e,\rm max}$ would necessitate a higher $B$ to simultaneously fit the X-ray data. This increased $B$ would, in turn, lead to a higher X-ray-to-TeV flux ratio (which scales as $\propto B^2$), rendering a simultaneous fit to both the X-ray and VHE flux levels impossible. Furthermore, high $B$ would also increase synchrotron cooling, which would reduce the TeV emission and broaden the X-ray spectrum.

Alternatively, the discrepancy might be addressed by including an additional seed-photon field or a separate electron population. If a plausible external seed-photon field (e.g., optical) is introduced, IC emission from $E_e\approx 50$\,TeV electrons might reproduce the TeV excess. We explored this possibility by introducing a seed-photon field with a temperature $kT=10^4$\,K and energy density $u_{\rm seed}=6000 \rm \ eV\ cm^{-3}$. While this field can help reproduce the observed flux of $\gapp 10^{-12}$\,\fluxcgs\ at $\lapp$TeV energies, the required energy density is unrealistically high. Moreover, the resulting SED shape is very broad, leading to the model significantly overpredicting the data at energies $>1$\,TeV. Furthermore, in this case the IC emission from the radio-emitting electrons significantly overpredicts the low-energy Fermi-LAT measurements at $\lapp 10$\,GeV. 

These considerations suggest that an additional population of electrons is needed to explain the $\lapp$TeV excess in the multi-zone framework. Although simply including an additional electron population can readily account for the $\lapp$TeV excess, the existence of such a population remains unjustified within the framework of our current multi-zone model. We explore the implications and possible origins of this additional particle population in Section~\ref{sec4_3}.

\section{Discussion}
\label{sec4}
We performed a comprehensive X-ray analysis of the PWN CTB 87, using new and archival data from the Chandra, XMM-Newton, and NuSTAR observatories. Our analysis provided detailed measurements of the broadband X-ray spectrum and the radial profiles of the source's surface brightness and $\Gamma$. We subsequently employed PWN models to explain our X-ray measurements and the broadband radio-to-TeV SED.

\subsection{X-ray Emission from CTB~87}
\label{sec4_1}
Our analysis of the X-ray spectrum, covering a wider energy range than previous studies, provided a more accurate characterization of CTB 87's emission. While our re-analysis of the Chandra and XMM-Newton data confirmed $\Gamma$ values of 1.7--1.9 \citep[][]{Matheson2013,Guest2020}, the new NuSTAR observations revealed a significantly softer $\Gamma$ of $2.74\pm 0.10$. This discrepancy in $\Gamma$ values cannot be attributed to residual pulsar contamination or to cross-calibration uncertainties \citep[e.g.,][]{Madsen2017}, suggesting a curved X-ray spectrum \citep[see][for other PWNe]{Bamba2022}.

A joint analysis of the combined Chandra, XMM-Newton, and NuSTAR data revealed highly significant spectral curvature in our BPL or CPL fits (Section~\ref{sec2_4_2}), where the CPL model yielded a smaller $\chi^2$ value than the BPL model. Furthermore, individual analyses of the XMM-Newton and NuSTAR data with a BPL indicate gradual curvature rather than a sharp break. Based on these findings, we favor the CPL model, which implies a hard electron injection with a PL index of $p_1 \sim 1.5$. Such a hard injection suggests that particle acceleration via magnetic reconnection is at play within this PWN \citep[e.g.,][]{Sironi2014}.

We further characterized the spatial distribution of the emission by measuring the X-ray surface brightness and $\Gamma$ profiles. Our analysis confirms the radial softening of the spectrum previously reported by \citet{Guest2020} and extends this finding to higher energies, providing strong evidence of the synchrotron cooling scenario.

\subsection{Evolutionary One-zone Modeling}
\label{sec4_2}

We interpreted the broadband emission of CTB 87 using an evolutionary one-zone model \citep[][]{Gelfand2009}. This model successfully reproduced the broadband SED (Figure~\ref{fig6}) with parameter values typical for a PWN-SNR system, notably suggesting that the PWN has been interacting with a reverse shock since $\sim 1.6$~kyr, with the compression peaking at $\sim$4.7\,kyr, followed by re-expansion to its present radius.
Our model results differ from those of previous stationary one-zone modeling \citep[][]{Xin2025,Gong2025} because the inclusion of the dynamic evolution of the PWN accounts for the 1--10\,GeV emission without invoking additional processes. 
Additionally, our refined X-ray measurements placed tighter constraints on the maximum electron energy.

However, this model faces significant limitations. 
The X-ray SED predicted by the one-zone model is flatter than our X-ray measurements (see inset of Figure~\ref{fig6}), such that the model overpredicts the $\sim 2$~keV X-ray flux. 
The synchrotron rollover of a single, smoothly cut-off electron population in a single magnetic field is broader than the observed $\sim$5\,keV break.
Similarly, the model predicts a broad and smooth VHE SED which does not account for the sharp SED feature at $\sim0.1\textrm{--}1$\,TeV energies. Furthermore, our observed X-ray spectral softening provides clear evidence of spatially varying PWN properties that one-zone models cannot accommodate. 

\subsection{Multi-zone Emission Modeling}
\label{sec4_3}
To address these discrepancies, we adopted a multi-zone model. We selected a relatively young age of $t_{\rm age} = 5$\,kyr, guided by the measured X-ray-to-TeV flux ratio and the point-like TeV morphology reported by VERITAS \citep[][]{Abeysekara2018}. Older ages would predict a lower flux ratio and a spatially extended TeV emission, both of which are incompatible with observations. Note that if higher TeV fluxes are confirmed by future measurements, such as those suggested by HAWC \citep{Albert2021}, the model could accommodate an older age.

The multi-zone framework successfully explains the observed X-ray spatial variations and SED. Specifically, the hard X-ray spectrum below $E_{\rm c} \approx 5$\,keV necessitates a hard power-law injection, while $\sim20$\,keV and the $\ge10$\,TeV data constrain the maximum electron energy to $E_{e, \rm max} \sim 200$\,TeV. 

Despite its successes, our multi-zone model underpredicts the Fermi-LAT SED in the 1\,GeV--1\,TeV range. The complex SED shape in this energy band suggests multiple emission components: a low-energy ($\lesssim 10$\,GeV) and a high-energy ($\sim 0.1\text{--}1$\,TeV) component. Previous studies have attributed the low-energy emission ($\lesssim 10$\,GeV) to a low-energy Maxwellian electron population \citep[][]{Saha2016} or pion decay in the SNR shock \citep[][]{Xin2025}. Alternatively, this could arise from the pulsar J2016 itself. Although its gamma-ray pulsations remain undetected, our modeling confirms that typical pulsar parameters (e.g., 4FGL catalog) could explain this flux component (green curve in Figure~\ref{fig7}). Detection of GeV pulsations by Fermi-LAT would definitively confirm the pulsar's contribution.

Given the spectrally hard electron distribution required by the observed narrow X-ray SED and hard $\Gamma$ in the inner PWN regions \citep[Figure~\ref{fig5}b; see also][]{Guest2020}, the broad VHE SED, especially the $\sim0.1\textrm{--}1$\,TeV excess, cannot be explained by our multi-zone model. This discrepancy requires an additional population of electrons that emit synchrotron and IC photons in the optical and $\sim0.1\textrm{--}1$\,TeV bands, respectively. A viable explanation involves an additional low-energy electron population in specific PWN zones (e.g., TeV emission regions), formed by reverse-shock compression \citep[e.g.,][]{Slane2010, Pope2024}. Alternatively, two distinct acceleration processes may operate within the PWN: Fermi-I acceleration in the equatorial region and magnetic reconnection in the bulk of the nebula, producing separate electron populations as proposed by \citet{Lyutikov2019}.

Although the observed $\sim0.1\textrm{--}1$\,TeV excess hints at the presence of a cooled population, its statistical significance is not yet high. Further observational confirmation is necessary; if verified, detailed modeling of this $\lapp\rm TeV$ feature can provide important insights into the evolutionary history of the PWN.

\subsection{Comparison Between the One-zone and Multi-zone Results}
\label{sec4_4}
While the one-zone and multi-zone models yield consistent SED shapes and parameter values for very young systems like G0.9$+$0.1 \citep[e.g.,][]{Brunelli2026}, their predictions for the more evolved PWN CTB~87 differ significantly. This discrepancy stems from the fundamentally distinct physical scenarios, data priorities, and evolutionary treatments (e.g., temporal vs spatial) adopted by each framework.

The primary difference lies in the treatment of the X-ray data and temporal evolution. 
The one-zone model fit the global broadband SED but did not fully reproduce the detailed X-ray spectral curvature.
Within this framework, incorporating dynamic evolutionary effects allows the model to reproduce the radio and 1--10\,GeV flux via reverse-shock compression.
Conversely, the multi-zone model prioritized the X-ray measurements, successfully reproducing the hard $\Gamma$ in the inner PWN and subsequent radial softening. However, this stationary multi-zone framework significantly underpredicts both the radio and sub-TeV data. While the radio and GeV deficits can be physically reconciled, by invoking a separate low-energy electron population and accounting for the putative gamma-ray contribution of J2016, the excess flux at $\lesssim$\,TeV energies may require an additional, cooled electron population, likely associated with reverse-shock compression.

These distinct emission scenarios naturally lead to some differences in the inferred/assumed physical parameters. For instance, the volume-averaged $B$ is higher in the multi-zone model ($\sim 13\,\mu$G) because it explicitly accounts for spatial transport. Particles escape the active volume via diffusion, and this particle loss was compensated for by a stronger $B$ to reproduce the observed X-ray flux. Consequently, the inferred $E_{e,\rm max}$ differs by a factor of $\sim 2$ between the two models. This is because the highest-energy synchrotron emission originates in the innermost region where the multi-zone local $B$ is strongest ($E_{e,\rm max} \propto B^{-1/2}$). However, given the strong parameter covariances in both models, these values should be interpreted as order-of-magnitude estimates, meaning the apparent numerical tension is not highly significant.

Because of these differing model assumptions, we did not enforce identical geometric or environmental parameters across both frameworks. The physical definitions of PWN size differ fundamentally; the one-zone model relies on a global, volume-averaged radius, whereas the multi-zone model prioritizes the distance from the pulsar to reproduce the observed radial profiles. Therefore, for an asymmetric PWN like CTB~87, the adopted $R_{\rm PWN}$ values can significantly vary between the two approaches. 
Similarly, the multi-zone model fixes parameters like $t_{\rm age}$ and $d$ based on independent empirical constraints (Section~\ref{sec3_2}), while the one-zone model optimizes them during the spectral fit ($d$ subject to a Gaussian prior centered on the same HI distance, and $t_{\rm age}$ derived from the fitted braking index and spin-down time scale together with the measured pulsar spin parameters).
Nonetheless, these values are not critically discrepant, and alternative values can easily be accommodated by adjusting other parameters.

In summary, while neither framework independently accounts for the detailed broadband data, both ultimately require multiple electron populations that may originate from evolutionary effects or distinct acceleration mechanisms. Observationally, these are demanded by the challenge of simultaneously modeling a narrow, highly curved X-ray SED alongside a broad, complex gamma-ray SED observed in CTB~87.

\section{Summary}
\label{sec5}
We have presented an analysis of the broadband X-ray spectrum of CTB 87, revealing significant spectral curvature well-described by a CPL ($\Gamma=1.23$, $E_{\rm cut} \approx 5.3$\,keV). This hard photon index implies a similarly hard injection spectrum of electrons, likely accelerated via magnetic reconnection.

While our evolutionary one-zone and multi-zone models successfully capture several observational features, they do not fully account for the complex SED, particularly the GeV-to-TeV emission components. Addressing these issues will require deep VHE observations to clarify the origin of the $\sim$\,TeV bump and potential detections of GeV pulsations to isolate the pulsar's contribution. Our findings suggest that CTB 87 is a complex, evolving system. Evolutionary multi-zone modeling will be essential to reconstruct the evolutionary history of this PWN.

\bigskip
\bigskip

\begin{acknowledgments}
We thank the anonymous referee for the constructive comments. This work used data from the NuSTAR mission, a project led by the California Institute of Technology, managed by the Jet Propulsion Laboratory, and funded by NASA. We made use of the NuSTAR Data Analysis Software (NuSTARDAS) jointly developed by the ASI Science Data Center (ASDC, Italy) and the California Institute of Technology (USA). This research was supported by the National Research Foundation of Korea (NRF) grant funded by the Korean Government (MSIT) (RS-2023-NR076359). C.K. acknowledges support from Basic Science
Research Program through the National Research Foundation of Korea (NRF) funded by the Ministry of Education RS-2025-25406195). S.S.H. acknowledges support from the Canadian Space Agency (CSA) and from the Natural Sciences and Engineering Research Council of Canada (NSERC) through the Canada Research Chairs and the Discovery Grants programs. This paper employs a list of Chandra datasets, obtained by the Chandra X-ray Observatory, contained in Chandra Data Collection\dataset[DOI: 10.25574/cdc.456]{https://doi.org/10.25574/cdc.456}.

\end{acknowledgments}

\bigskip

\vspace{5mm}
\facilities{CXO, XMM, NuSTAR}
\software{HEAsoft \citep[v6.34;][]{heasarc2014}, CIAO \citep[v4.16;][]{CIAO2013}, XMM-SAS \citep[v22.1.0;][]{xmmsas14}, XSPEC \citep[v12.14.0;][]{a96}}

\bibliography{ctb87}
\bibliographystyle{aasjournalv7}
\end{document}